# Realization of one-dimensional electronic flat bands in an untwisted moiré superlattice

Yafei Li[1,†], Qing Yuan[1,†], Deping Guo[2,†], Cancan Lou[1], Xingxia Cui[1], Guangqiang Mei[1], Hrvoje Petek[3], Limin Cao[1], Wei Ji[2,*] & Min Feng[1,4,*]

[1]School of Physics and Technology and Key Laboratory of Artificial Micro- and Nano-Structures of Ministry of Education, Wuhan University, Wuhan 430072, China
[2]Beijing Key Laboratory of Optoelectronic Functional Materials & Micro-Nano Devices, Department of Physics, Renmin University of China, Beijing 100872, China
[3]Department of Physics and Astronomy and IQ Center, University of Pittsburgh, Pittsburgh, PA 15260, USA
[4]Institute for Advanced Study, Wuhan University, Wuhan 430072, China

†These authors contributed equally to this work.
*Corresponding author. Email: wji@ruc.edu.cn (W.J.); fengmin@whu.edu.cn (M.F.)

**Abstract: Two-dimensional electronic flat bands and their induced correlated electronic interactions have been discovered, probed, and tuned in interlayer regions of hexagonally shaped van der Waals moiré superlattices. Fabrication of anisotropic one-dimensional correlated bands by moiré interference of 2D, however, remains a challenge. Here, we report an experimental discovery of 1D electronic flat bands near the Fermi level in an anisotropic rectangular moiré superlattice composed of in situ grown, vdW stacked two-atomic-layer thick Bi(110) well-aligned on a SnSe(001) substrate. The epitaxial lattice mismatch between the aligned Bi and SnSe zigzag atomic chains causes strong three-dimensional anisotropic atomic relaxations with associated one-dimensional out-of- and in-plane strain distributions that are expressed in electronic bands of the Bi(110) layer, which are characterized jointly by scanning probe microscopy and density functional theory. At the regions of the strongest out-of-plane shear strain, a series of 1D flat bands near the Fermi level are experimentally observed and defined in our calculations. We establish that 1D flat bands can arise in moiré superlattices in absence of the relative layer twist, but solely through the lattice strain. We generalize the strategy of utilizing strain in lattice mismatched rectangular hetero-bilayers for engineering correlated anisotropic electronic bands.**

Correlated electronic quantum phenomena in complex and/or low dimensional materials are of fundamental interest[1-5] and technological importance[6-12] in discovery of novel quantum materials. Homo and heterobilayers of graphene, transition-metal dichalcogenides (TMD), and related materials have been stacked at defined twisting angles[2,11,13-23] or with a small lattice mismatch[24,25] to form two-dimensional (2D) moiré superlattices with underlying hexagonal/triangular symmetry to create band interference effects to subtly tune the in-plane kinetic energy scales and thereby enhance the interactions among the carriers[2-4,13-24,26-29]. They are thus attractive platforms for designing and exploring collective quantum states of matter, such as, correlated insulators[2,4,19-21], Wigner crystals[3], pair density waves in high-temperature superconductors[13,14,18,26], with an unprecedented potential for the design and control[2,15-17,22-24,26].

In addition to achieve 2D isotropic flat bands with the approach of control of the relative twist angle between combined sheets, recently, it has been predicted that the twisted rectangular homo-bilayers of GeSe[30] or SnS[31] could also harbor one-dimensional (1D) flat bands. Such 1D flat bands have been predicted to harbor anisotropic correlated quantum phenomena[30-32], such as Majorana edge states[27], promoting exploration of their novel electronic properties as exemplars of emergent solid-state quantum materials. Although numerious twisted hexagonal/triangular bi-layers have been achieved, controllable composition of rectangular bilayers with defined twisting angle has not been reported. Thus, the theoretically predicted twisted GeSe or SnS homo-bilayers and other twisted rectangular superlattices are yet to be experimentally realized. Alternatively, considering the technological challenge in delicately controlling the twisting angles between atomic layers, a more feasible approach may be to grow untwisted moiré superlattices that intrinsically host 1D flat bands.

Here, we report experimental realization of 1D electronic flat bands in an untwisted, but lattice-mismatched hetero-bilayer. We grow a rectangular moiré superlattice by vdW epitaxy of a two atomic-layer (2-AL) Bi(110)[33-35] on a SnSe(001) substrate. The anisotropic substrate lattice has relatively small mismatch along zig-zag (ZZ) direction, and substantially larger one along arm-chair (AC) directions. This causes the Bi layer growth to form a rectangular moiré superlattice, even in the absence of a twist angle. By density functional theory (DFT) calculations and q-Plus non-contact atomic force microscopy (q-Plus nc-AFM) measurements, we characterized substantial 3D atomic relaxations of the Bi(110) layer, which portend spatially-locked in-plane tension and out-of-plane shear that are distributed as 1D strain stripes in the $ZZ$ direction. Scanning tunneling microscopy/spectroscopy (STM/STS), together with DFT calculations, reveal several 1D electronic flat-bands around the Fermi level ($E_F$), where electrons are delocalized with only several meV dispersion along the $ZZ$ direction and essentially localized with no dispersion along the $AC$ direction. Moreover, the spatial distribution of the wavefunction norms (($|\psi|^2$) of these flat bands nearly superpose over the strain stripes. This superposition is not a coincidence, but reveals a profound relationship between the strain and the flat band. Given the correspondence between the lattice-mismatch induced strain stripes and the 1D flat bands, our work establishes a promising novel strategy for designing scalable 1D quantum systems in untwisted 2D moiré superlattices.

## Results

**Atomic structure of the Bi(110) superlattice.** Figure 1a shows an atomic model of a 2-AL Bi(110) layer residing on a SnSe(001) substrate. The 2-AL Bi(110) layer, a structural analogue of monolayer black phosphorous, is composed of Bi atoms propagating in chains along the $ZZ$ direction that form armchair-like ridges and troughs in the orthogonal the $AC$ direction (Fig. 1b)[36-38]. The SnSe(001) substrate shares the same atomic structure except that the Sn and Se atoms in the same atomic layer are slightly buckled (Fig. 1a)[39-41]. STM images of the as-grown Bi(110) film (Fig. 1c) show waffle-like rectangular superlattice surface structures that exhibit different contrast distributions under acquisition with positive (0.40 V, Fig. 1d) and negative (-0.10 V, Fig. 1e) bias voltages. According to the STM contrast and STS spectra, to be discussed later, we divide the supercell into four regions (marked with "1" to "4" in Figs. 1d and 1e); their imaging under different bias voltages is presented in Fig. S1. The size of each supercell is ~53.7 Å ($AC$) × 50.1 Å ($ZZ$) corresponding to a 11 × 11 Bi(110) unit cell with the measured Bi(110) lattice constants of 4.8 ± 0.2 ($AC$) and 4.5 ± 0.2 ($ZZ$) Å. Figure 1f shows the atom-resolved STM image of the 11 × 11 Bi(110) unit cell and it four regions. An STM image taken at an edge of the Bi(110) island (Fig. S2) confirms that this 11 × 11 superlattice is well aligned with the $ZZ$ and $AC$ directions of the SnSe(001) substrate, and therefore in not twisted. Given the observed alignment of the superlattice and the measured SnSe(001) lattice constants of $4.5 \pm 0.2$ ($AC$) and $4.2 \pm 0.2$ ($ZZ$) Å, we conclude that Bi forms a Bi(110)-11×11/Sn(001)-12×12 moiré superlattice. The 12×12 SnSe substrate shows lattice mismatches of +2.27% ($AC$) and +1.81% ($ZZ$) to the Bi layer, implying an anisotropic 2D lattice tension to exist in the 2-AL Bi.

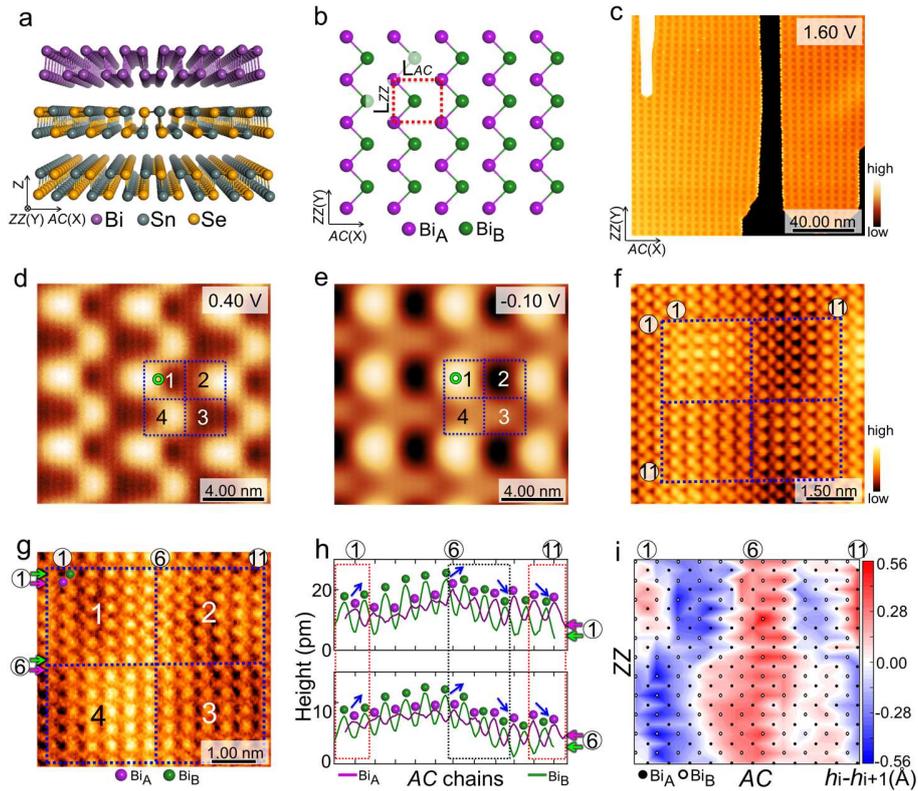

**Fig. 1 2-AL Bi(110) superlattice on SnSe(001). a** Side view of a structural model of a well aligned Bi/SnSe interface. **b** Top view of an isolated 2-AL Bi(110) layer with only the top atomic layer Bi atoms, $Bi_A$ and $Bi_B$, shown, where the red-dotted rectangle defines the unstrained layer unit cell. **c** STM image of Bi islands grown on the SnSe substrate. **d, e** Two typical STM images acquired at +0.40 V (I = 20 pA) and -0.10 V (I = 20 pA), respectively. The blue-dotted rectangle represents the supercell composed by the four contrast regions "1" to "4". The green circled-dot marks the same real space locations in (**d**) and (**e**), indicating a contrast shift. **f** Atomic resolution STM image of the Bi(110) supercell (0.14 V and 100 pA) with the blue dotted rectangle marked the same supercells as in (**d**) and (**e**). The 1[st] and 11[th] chains/rows are marked. **g** Atomic resolution constant $\Delta f$ mode q-plus AFM image (CO tip, $\Delta f$ = -3.85 Hz) with the $Bi_A$ and $Bi_B$ sites marked. Other markers are the same as those in (**f**). **h** Line profiles cutting along $Bi_A$ and $Bi_B$ atoms [marked (**g**)] in the 1[st], 6[th] $AC$ rows. The blue arrow marks the buckling direction from $Bi_A$ to $Bi_B$. which shows a reversion in the locations marked by the red and black dotted rectangles. **i** The $h_i$-$h_{i+1}$ map based on the $h_i$ in (g). $Bi_A$ ($Bi_B$) is represented by black dot (circle).

Given the sharp STM contrast of the supercell under positive (Fig. 1d) and negative (Fig. 1e) bias, the surface is next characterized by q-plus nc-AFM. Figure 1g shows a typical atom-resolved nc-AFM image, where we locate regions 1-4 of the STM contrast within the AFM image after exacting comparison of the AFM and STM images (Fig. S3). The AFM image contrast records the height variation of the top atomic layer (T-AL) Bi atoms. It reveals a pronounced site-specific relaxation of individual Bi atoms in the 2-AL Bi layer. This is illustrated by taking line-profiles along the $AC$ direction through Fig. 1g, which are displayed in Fig. 1h. The two Bi atoms of the unstrained atomic unit cell are labeled as $Bi_A$ and $Bi_B$ in Fig. 1b. The selected $AC$ atomic chains, i. e., the 1st, 6th chains, exhibit typical pronounced buckling of the neighboring Bi atoms, which importantly, reverses around the inter-supercell (marked by red-dotted rectangle) and intra-supercell boundaries (black-dotted rectangles), respectively. This demonstrates that abrupt height dislocations occur at the boundary regions. Consequently, the supercell experiences sharp periodic out-of-plane shear-strain along the $AC$ direction, as depicted by a $\Delta h_i$ map in Fig. 1h ($\Delta h_i = h_i - h_{i+1}$, where $h_i$ and $h_{i+1}$ stand for the height of the adjacent Bi atoms in the atomic chains along $AC$ direction). The height analysis demonstrates that shear strain dominantly occurs at the Region boundaries and extends along the $ZZ$ direction. We note that this atomically resolved map directly records the atomic relaxation in moiré superlattice, whereas previous studies claimed the strain to exist based on an indirect proof[24].

**1D flat bands observable in the strained region.** Studies have shown atomic relaxation in moiré superlattices can profoundly affect on the electronic properties of moiré flat bands[24,42]. Our STS measurements show that the Bi(110) rectangular moiré superlattice hosts 1D flat electronic bands, which are closely related with the mapped shear strain. Fig. 2a presents typical STS d$I$/d$V$ spectra acquired from the centers of the four regions in the Bi(110) supercell. A common feature is that all the spectra host pronounced occupied state STS peaks, which are extremely sharp. Normally, a sharp STS peak implies the existence and defines the energy of an electronic flat band[24,42,43]. The characteristic sharp STS peak is located at ~-0.14 V in the spectrum from Region 1, while in Regions 2-4, such peaks appear at ~-0.27 V to ~-0.30 V, respectively. Because the STS maps at ~-0.27 V to ~-0.30

V are essentially the same (Fig. S4), here we mainly focus on characteristics of the -0.14 and -0.30 eV states acquired from Region 1 and 2.

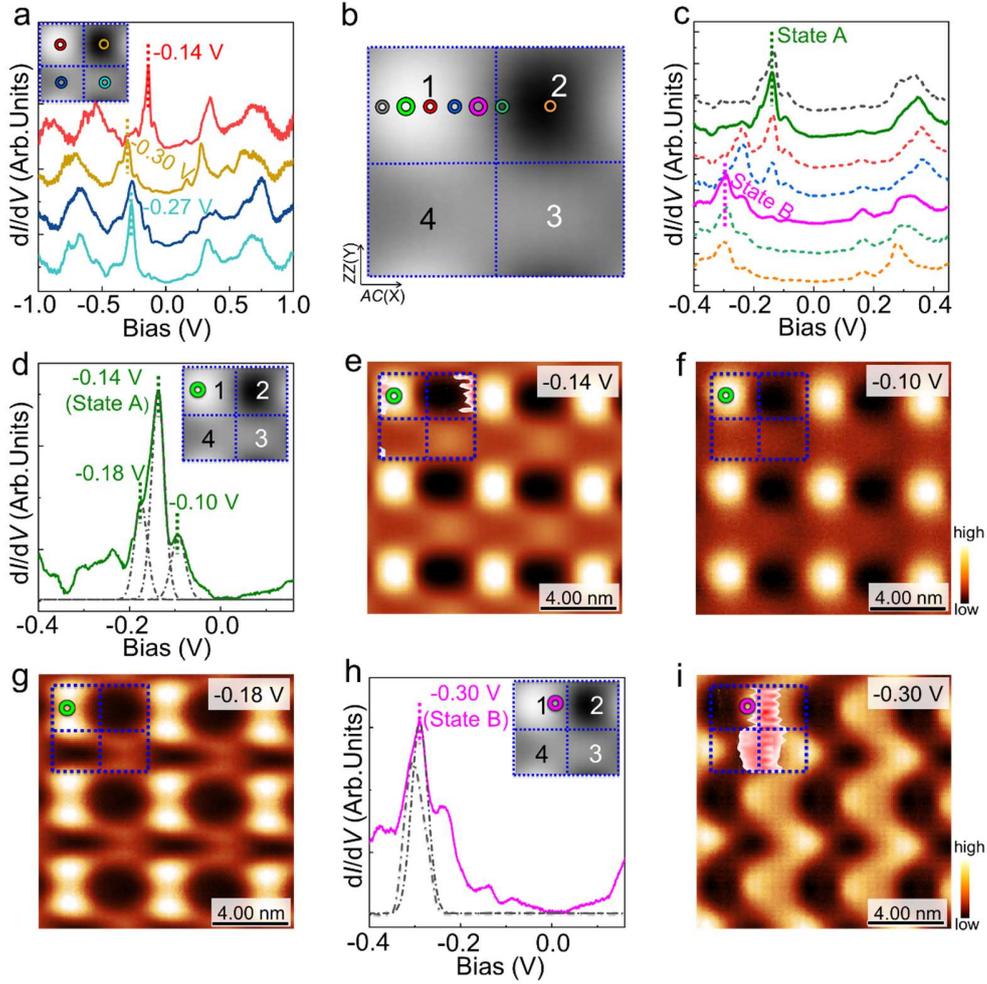

**Fig. 2 Experimentally established 1D flat band electronic states in 2-AL Bi(110) superlattice on SnSe(001). a** Typical STS d$I$/d$V$ spectra acquired around the center of Regions 1-4 (inset). **b, c** site-dependent STS spectra (**c**) measured at the circle marked locations in (**b**). States A and B are highlighted. **d** Zoomed-in STS spectrum recorded at the green circle location (inset) with the analysis of the -0.14 V peak. **e** to **g** STS d$I$/d$V$ maps obtained at the energies of -0.14, -0.10, and -0.18 V spectral components in (**d**), respectively. The blue dotted rectangles mark one supercell. **h** Zoomed-in STS d$I$/d$V$ spectrum acquired from the pink circle location (inset) with the analysis of the -0.30 V peak. **i** STS d$I$/d$V$ map obtained at -0.30 V from (**h**). The dominant experimental shear strain distributions are overlapped in (**e**) and (**i**).

Site-dependent STS d$I$/d$V$ spectra (Fig. 2c) that are acquired at positions marked by the colored spots in Fig. 2b along the line bisecting Region 1 and spanning to the center of Region 2, show that the -0.14 V peak (labeled State A) gradually evolves into the -0.30 V peak (State B) when translating the measurement position along the *AC* direction. Moreover, the State A has the highest intensity at the green spot marked location in Region 1, while State B is most prominent at the pink spot site between Regions 1 and 2.

Figure 2d shows a zoomed-in d*I*/d*V* spectrum acquired at the green-spot position (replotted from Fig. 2b in the inset). The prominently narrow full width at half maximum (FWHM) of 29.5 mV for the -0.14 V State A approaches the instrumental resolution of 25 meV at T=5.0 K, which is expected from a flat band[24,42,43]. The LDOS map at this energy (Fig. 2e) shows nearly isolated bright domains dominantly located on the left side of Region 1, resembling the inter-boundary shear strain regions (semi-transparently superposed in Fig. 2e). The bright contrast of State A indicates its strong spatial localization, as is observed for localized moiré flat bands in TMD moiré superlattices[24,42,43]. In addition, the -0.14 V peak has several "satellite" fringe peaks shown in Fig. 2d. At its right, a peak residing at -0.10 V shows a comparable LDOS distribution to the -0.14 V peak (Fig. 2f). The satellite peak -0.18 V, however, has a "*dumb bell*-like'' pattern in its d*I*/d*V* map with a minimum intensity at the green spot (Fig. 2g) where the -0.14 V peak has its maximum intensity. These observed LDOS distributions are reminiscent of wavefunction distributions of the effective moiré potential quantum series responsible for the moiré flat bands in TMD superlattices[23,24,42,43]. However, contrary to the isotropic LDOS patterns of the flat bands in hexagonally shaped van der Waals moiré superlattices, the LDOS distribution of the -0.14 V State A is highly asymmetric: in the space between Region 1, the intensities are higher along *ZZ* direction than that along *AC* direction, indicating an anisotropic character of State A.

The -0.30 V State B possesses a more defined 1D flat band character. Again, the very narrow FWHM of 30.3 mV of its d*I*/d*V* peak suggests a flat band electronic character (Fig. 2h). The corresponding STS map, as shown in Fig. 2i, demonstrates that it is delocalized specifically along the *ZZ* direction forming a 1D stripe pattern, dominantly distributed on the intra-boundary strain regions mapped semi-transparently on Fig. 2i. Such striped LDOS for a 1D flat band that we reveal in an untwisted Bi/SnSe bilayer resembles similar strip-like feature that was predicted in twisted rectangular-lattice GeSe homo-bilayers[30]. Therefore, our STM experiments find that both States A and B possess most of the characteristic features that have previously been predicted in 1D electronic flat bands of twisted bilayers[30,31]. Moreover, we characterize how their spatial distributions are defined by the dominant shear strain regions.

**DFT analysis of the 3D atomic relaxations.** Our DFT calculations reproduce the experimentally observed 3D atomic relaxations in the Bi(110) layer and furthermore, confirm the strain origin of the 1D flat band electronic states. Figure 3a shows the fully relaxed atomic structures of the Bi(110)-11×11/SnSe(001)-12×12 hetero-interface, which exhibits a rectangular moiré pattern that contains four different domains (denoted I to IV; corresponding to experimental Regions 1-4), which primarily represent the four stacking orders shown in Fig. 3b. Figure 3c presents the calculated relaxed $h_i$ map of the Bi atoms in the top atomic layer (T-AL). As the Bi atoms in the bottom AL follow the Bi atoms in T-AL, exhibiting the same height trend, Fig. 3c thus represents the theoretical height map of the relaxed 2-AL Bi layer. The $h_i$ map shows pronounced height difference for the $Bi_A$ (black dots) and $Bi_B$ (black circles) atoms in one unit cell indicated by the white (high) and grey (low) contrast.

Similar to the experimental observations, variations in $h_i$ modulate the Bi $ZZ$ chains. $Bi_A$ and $Bi_B$ are thus buckled as noted in the side view of the 1st atomic $AC$ chain in the supercell (Fig. 3a lower panel), accompanied with a reversal buckling at the region boundaries. Consequently, abrupt height changes induce the out-of-plane shear-strain (represented as $h_i$ - $h_{i+1}$) along the $AC$ direction, as depicted in Fig. 3d (plotted using $h_i$ of $Bi_A$ atoms in Fig 3c). The shear-strain distribution exhibits two 1D stripe-shape wings extending along the $ZZ$ direction, which meet in Region I and follow the transition regions of II+IV. A similar shaped shear strain map is found for the $Bi_B$ atoms in the top AL (Fig. S5).

The out-of-plane shear strain is accompanied by in-plane tension and compression strains (the in-plane strain $\varepsilon = \frac{d-d_{av}}{d_{av}} \cdot 100\%$ is calculated by comparing the distance $d$ between each Bi atom and the nearest Bi atom along the $ZZ$ or $AC$ direction with the average lattice constant $d_{av}$. Figure 3e shows a 2D map of the in-plane tension (red and yellow) and compression (blue) distribution in the $AC$ direction. It is notable that the in-plane $AC$ tension spatially coincides with the shear strain shown in Fig. 3d. Along $ZZ$ direction, the in-plane compressive strain (blue) is observed where a tensile strain (red and yellow) along AC occurs, which is illustrated in Fig. 3f. Comparing the scales of the color bars in Fig. 3e and 3f shows that the highest tension and compression along the $AC$ direction are roughly three times higher than along the $ZZ$ direction. This ratio is consistent with the previously reported theoretical Poisson ratio $v_{x(AC)y(ZZ)} = 0.333$[44] for a 2-AL Bi(110), indicating the dominant strain exists along the $AC$ direction, consistent with the larger mismatch between the Bi and SnSe lattices in this direction.

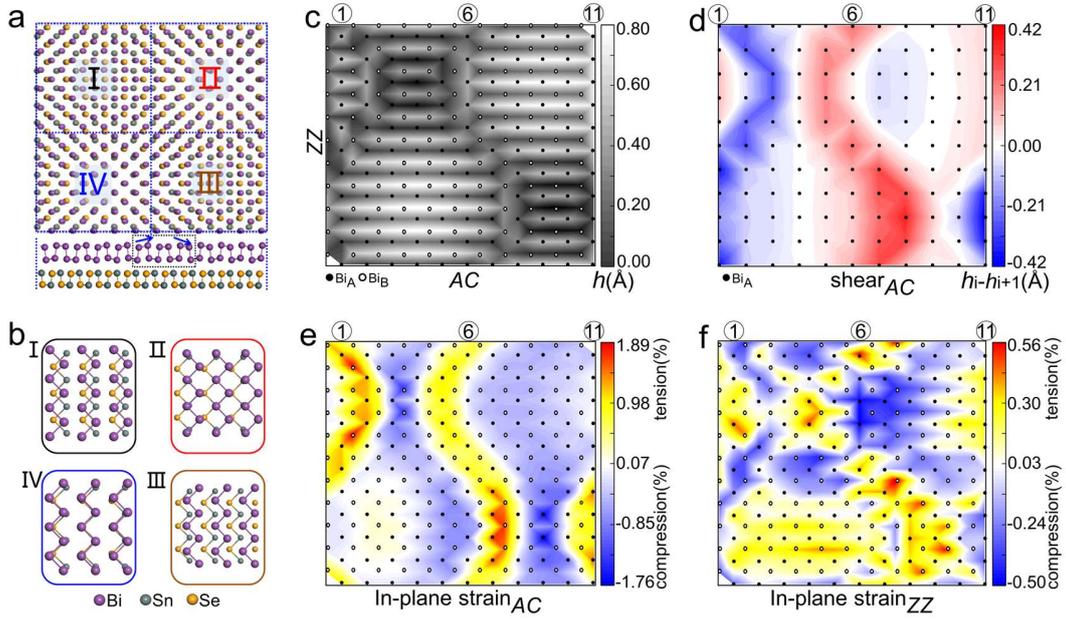

**Fig. 3 Theoretically quantified 1D strain fields in Bi(110) moiré superlattice on SnSe(001). a** Atomic model of the relaxed $11L_{AC} \times 11L_{ZZ}$ 2-AL Bi(110) on SnSe(001), characterized as four domains marked by "I" to "IV". The black dotted rectangles in the side view image of the 1st arm-chair chains mark the location with opposite Bi atom buckling happening, shown by the blue arrows. **b** Stacking configurations at the four domain centers. **c** Calculated 2D $h_i$ map of the top layer Bi atoms with black dots (circles) identifying the $Bi_A$ ($Bi_B$) atoms in one supercell. 1st, 6th,11th chains are marked. **d** 2D $\Delta h_i$ map representing shear strains experienced by $Bi_A$ atoms along $AC$ direction in the supercell, which are at the domain boundaries (around 1st, 6th and 11th chains). **e, f** In-plain strain (tension and compression) mapping along the $AC$ direction (**e**) and $ZZ$ direction (**f**), respectively.

**Strain induced 1D flat bands.** Our DFT results further reveal that the strained 2-AL Bi layer contains two 1D electronic flat bands as a consequence of the 3D atomic relaxation. Figures 4a and 4b show the band structures of a strain-free and a strained 2-AL Bi layers, respectively. For the strained case, the Bi(110) supercell is first relaxed on SnSe(001) substrate, and then the substrate is removed to calculate the band structure the strained bare Bi(110) supercell. The stars in Fig. 4a and 4b mark the folded conduction (CB) and valence bands (VB) around the Γ point, which are mainly comprised of the Bi $p_z$ orbitals (Fig. S6). The same colors connect related bands in both band structures while the numbers denote the band degeneracy. Both VBs and CBs of the strain-free 2-AL Bi layer are highly dispersive along both the Γ–X and Γ–Y directions (Fig. 4a), but those bands are strongly affected by strain (Fig. 4b).

In the strained Bi layer (Fig. 4b), the energies of all four $p_z$ bands of interest shift toward the Fermi level and those bands marked by the pink and green stars break from four-fold degeneracy into pairs of two-fold degenerate ones. All those bands have extremely narrow bandwidths, < 0.5 meV, along the Γ–X direction [the *AC* direction of Bi(110)]. Along the Γ–Y (*ZZ*) direction they are rather dispersive showing bandwidths varying from several to ~ 40.0 meV (Fig. S7). Two exceptions are found for two-fold degenerate bands sitting at 0.045 (green shading in Fig. 4b) and -0.041 eV (pink shading), respectively, which, however, exhibit extremely small bandwidths of 9.1 meV (Fig. 4c) and 3.2 meV (Fig. 4d) along the Γ–Y direction. According to few and nearly zero meV dispersions in the Γ–Y and Γ–X directions, as those predicted in the twisted GeSe homo-bilayers [30,31], these two bands are two 1D electronic flat bands.

The spatial distributions of the wave function norm square ($|\psi|^2$) of bands at 0.045 eV (Fig. 4e) and -0.041 eV (Fig. 4f) at the Γ point exhibit a 1D stripe shape extending along *ZZ* direction, which are correlated with and qualitatively follow the shear strain distribution: the $|\psi|^2$ of the 0.045 (-0.041) eV band follows the left (right) wing of the shear strain (the strain distributions are decorated with yellow-dotted stripe). These results explicitly show that electrons in these two bands are electronically localized in space along the *AC* direction and simultaneously delocalized in the *ZZ* direction with extremely small dispersions, giving them 1D flat band character.

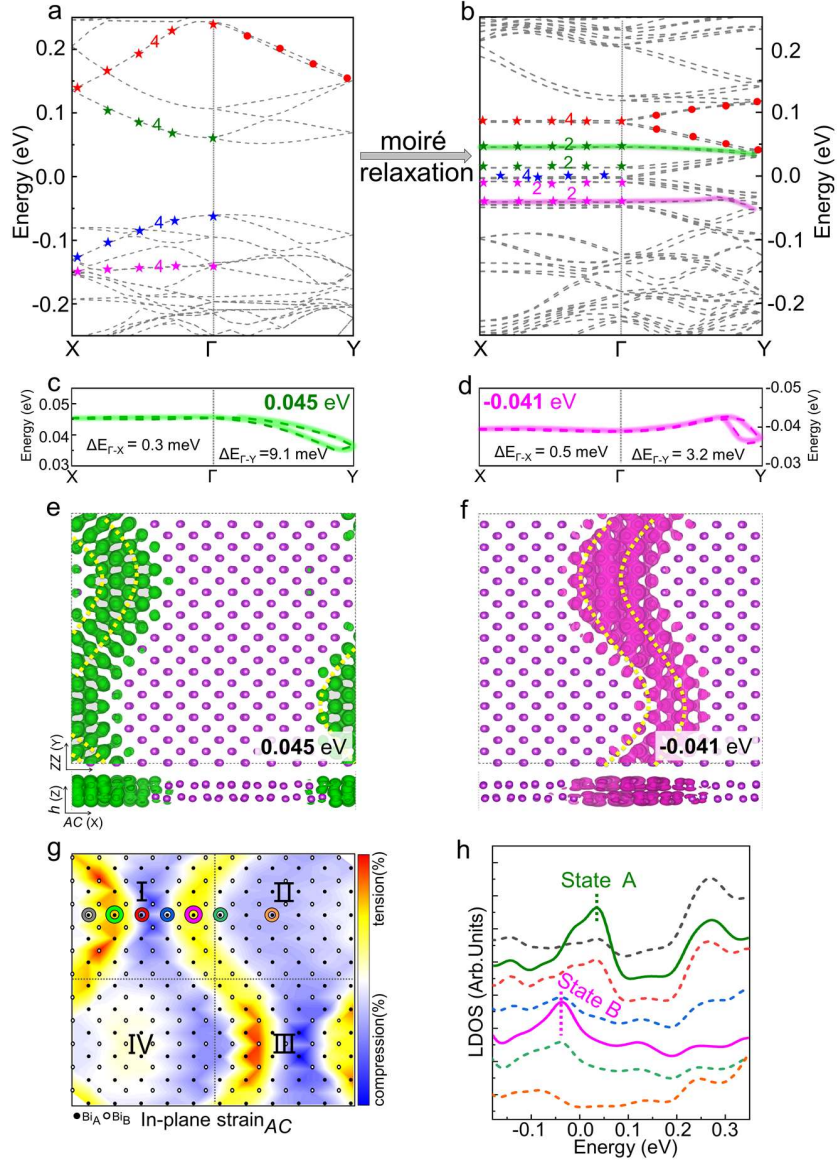

**Fig. 4 Strain induced 1D flat bands in 2-AL Bi(110) superlattice on SnSe(001). a, b** Band structures of $11L_{AC} \times 11L_{ZZ}$ 2-AL Bi(110) without (**a**) and with (**b**) 3D structure strain. The colored stars mark the related bands in (**a**) and (**b**). The green and pink shadowing in (**b**) mark two 1D flat bands, at +0.045 eV and -0.041 eV respectively. **c, d** Expanded images showing the bandwidths along X-Γ-Y path for the two highlighted 1D electronic flat bands in (**b**). **e, f** Top and side views of the $|\psi|^2$ at the energies corresponding to the two 1D electronic flat bands in (**b**). The yellow-dots strips in (**e**) and (**f**) highlight the locations of the 1D strain fields. The isosurfaces are $2 \times 10^{-5}$ e/Bohr³. **g** The in-plane strain fields along $AC$ direction with circles marked the locations for calculating the LDOS in **h**, where States A and B are assigned.

The 1D flat band characters of these two bands stand out as sharp peaks in the calculated LDOS, corresponding to the observed States A and B in the STS measurements.

This is verified by the calculated site-dependent LDOS, which are plotted along the line across the center location where the two strain stripes met in Domain I (Fig. 4g and 4h). The results are remarkably consistent with the measured LDOS of States A and B. Specifically, the two major peaks at 0.045 eV and -0.041 eV in Fig. 4h have similar relative energies and spatial distributions as the experimental ones. In particular, the 0.045 eV peak centered at the green-spot, representing the left wing of the stripe-strain regions, gradually losses its intensity along the $AC$ direction. Simultaneously, another pronounced peak at -0.041 eV evolves around the pink-spot, representing the right wing of the stripe-strain regions. Such experiment (Fig. 2b, c)-theory (Fig. 4g, h) consistency on these location-dependent spectra verifies our assignment of the 0.045 and -0.041 eV related flat bands to the experimentally observed States A and B respectively.

A remaining difference between the experiment and theory is the difference between the energies of States A and B. Considering that the calculated LDOS (Fig 4h) confirms the main characteristics of the measured STS spectra (Fig. 2c), the energy difference can be attributed mostly to the fact that band structure in Fig. 4b is for a freestanding strained Bi(110) sample. This is verified by finding that both the 0.045 eV and -0.041 eV bands shift to lower energies when calculated on the SnSe substrate (Fig. S8). One further difference between the experiment and theory is that the measurements are performed for a p-doped SnSe substrate (Fig. S9), while the calculation is conducted for a neutral one. The important finding, however, is that the experimentally and theoretically established 1D flat bands are robust features that are marginally influenced by electronic interactions with substrate.

**Discussions** The established direct correspondence between strain distributions and flat band state formation is further verified by finding that, like the State A and B, the $|\psi|^2$ distributions of other $p_z$ flat bands at $\Gamma$ point in Fig. 4b are also localized in either locations of the two strains (Fig. S7). For comparison, band structures of 2-AL Bi(110) on SnSe without relaxation (Fig. S10) are calculated and no flat band electronic states appear, unambiguously confirming that the 3D atomic relaxation caused by the moiré superlattice is responsible for the 1D flat bands. As the strain stripes are isolated along $AC$ ($\Gamma$–X) direction, we propose that the electron localization along this direction results from the trivial strain-potential induced spatial isolation of VBs and CBs. The limited electron localization along $ZZ$ (Γ–Y) direction, however, seems not trivial but originated from a fundamental electronic reason related with mirror (reversal) buckling of the $Bi_A$ and $Bi_B$ in $ZZ$ atomic chains. Further explorations are undertaken.

In summary, by utilizing vdW epitaxial growth, we build an aligned lattice-mismatched moiré superlattice, which intrinsically hosts 1D electronic flat bands localized by its out-of-plane strained regions. This strain-assisted strategy for self-assembled untwisted bilayers, without the need for demanding delicate control of twist angles between the vdW layers, simplifies the process of fabricating moiré superlattices that potentially host correlated electronic states. The same strategy can likely be extended to other group IV-VI phosphorene analogues[30,31], for which intrinsic anisotropic 1D moiré excitons[31] and Majorana zero modes[30] were predicted in twisted nanostructures. This highlights the significance of our material realization of the 1D electron flat bands as an important progress toward discovering anisotropic correlated 1D electronic states in 2D materials[32]. By achieving the microscopic subatomic resolution imaging of the moiré relaxation and the electronic characterization of the resulting flat bands, the discovered correspondence between the 1D stripe-regions under strain and the 1D electronic flat bands they support offers a strategy for the discovery and exploration of correlated electronic states in model moiré superlattices.

**Methods**

**Sample preparation.** The high-quality SnSe single crystals used in the experiments are self-grown using the temperature gradient growth method. High purity (99.9999%) Sn and Se granules are used for the sample growth. First, Sn and Se granules with the stoichiometry of SnSe and a total weight of 30 g are loaded into a quartz ampoule with the inner diameter of 11 mm. Then the ampoule is evacuated to better than $5\times10^{-5}$ Torr and sealed. Another quartz tube is evacuated and sealed to protect the sample and ampoule. The double-sealed quartz tube is loaded into a tubular furnace which is tilted 15° from the horizontal plane. The sample in the furnace is slowly heated to 980 °C over 30 hours, soaking at this temperature for 48 hours, and then cooling from 980 to 500 °C with a precisely controlled cooling rate of 1 °C h$^{-1}$. After cooling the furnace to room temperature, the synthesized SnSe single crystals with 5-10 mm width and/or length are taken out from the quartz ampoule into air.

Prior to the Bi growth and STM experiments, the SnSe crystals are cleaved *in-situ* in a preparation chamber under ultrahigh vacuum (UHV) at room temperature (RT). Bismuth atoms (99.999% purity, Aldrich) are evaporated from a resistively heated evaporator onto freshly cleaved SnSe surface. The SnSe substrates are kept at RT during the evaporation. The prepared sample is immediately transferred under UHV into the STM chamber, and cooled to 5.0 K.

**STM measurements.** The STM and spectroscopy experiments are carried out in a UHV low temperature STM system (CreaTec). STM topographic images are acquired in constant-current mode. The d$I$/d$V$ spectra are measured using the standard lock-in technique with a bias modulation of 8 mV at 321.333 Hz. The STM tips are chemically etched tungsten, which are further calibrated spectroscopically against the Shockley surface states of cleaned Cu(111) or Au(111) surfaces before performing measurements on Bi islands/SnSe. For the Q-plus AFM measurement, the tip is modified with CO by picking up a single carbon monoxide molecule from an Ag(100) surface. The parameters for picking up CO are sample bias of $V_b$ = 40 mV with tunneling current I = 100 pA. The AFM imaging is performed by frequency modulation (FM-AFM) with a constant amplitude of A = 120 pm. The resonance frequency of the AFM probe is $f_0$ = 24.5 kHz and the quality factor of Q is 53764.

**DFT Calculations.** DFT calculations are performed using the generalised gradient approximation for the exchange-correlation potential with a plane-wave basis and the projector augmented wave method as implemented in the Vienna ab-initio simulation package (VASP)[45-47]. The supercell for modeling of Bi/SnSe interface is composed of a 11 ×11 2-AL Bi(110) and a 12×12 2-AL SnSe layer, totally containing 1060 atoms. A 15 Å vacuum layer is adopted to avoid the image interactions. The Γ point is used for sampling the first Brillouin zone in structural relaxation and 3×3×1 in density of states (DOS) calculation. The energy cutoff for the structural relaxation and electronic structure calculations of the superlattice is set to 200 eV. The bottom SnSe layer is kept fixed and all other atoms are fully relaxed until the residual force per atom is less than 0.05 eV/Å during the optimization of the Bi-SnSe superlattice. In structural relaxation and electronic property calculations, the Grimme's D3 form vdW correction is applied with the Perdew–Burke–Ernzerhof (PBE) exchange functional (PBE-D3)[48,49]. All electronic properties of the superlattice are calculated and optimized with the consideration of spin-orbit coupling (SOC). For the in-plane strain ($\varepsilon = \frac{d-d_{av}}{d_{av}} \cdot 100\%$), the average calculated lattice constants ($d_{av}$) along the *AC* and *ZZ* directions are 4.87 and 4.45 Å, respectively, consistent with the experimental values. The height ($h_i$) mapping is calculated by subtracting the lowest height of Bi top layer from the real heights of the surface Bi atoms.

**Data availability:** The data that support the findings of this study are available from the corresponding authors upon request.

**Acknowledgements:** We gratefully acknowledge fruitful discussions with Chunwei Lin. This work is supported by the National Key R&D Program of China (Grant Nos. 2018YFE0202700, 2017YFA0303500, 2017YFA0303504), the Strategic Priority Research Program of Chinese Academy of Sciences (Grant No. XD30000000), the National Natural Science Foundation of China (Grant Nos. 11774267, 61674171, 11974422 and 11904349), and the Fundamental Research Funds for the Central

Universities and the Research Funds of Renmin University of China [Grants Nos. 22XNKJ30 (W.J.), and No. 21XNH090 (D.P.G.)]. H. Petek thanks partial support the Luo Jia Visiting Chair Professorship at Wuhan University, and NSF CHE-2102601 grant. Calculations were performed at the Physics Lab of High-Performance Computing of Renmin University of China.

**Author contributions**

Y.F.L. and Q.Y. grew the Bi samples and performed STM measurements. D.P.G. and W.J. performed first-principles calculations. C.C.L. and X.X.C. took part in the STM measurements. G.Q.M. and L.M.C. grew the SnSe single crystals. M.F. and L.M.C. initiated the project and experiments. W.J. conceived the theoretical calculations and analysis. F.C.W. participated in discussions. M.F., W.J., H.P. and L.M.C. analyzed the data, and wrote the manuscript with input from all authors.

**Additional information**

Additional data related to this paper are available from the corresponding authors upon request.

**Competing interests**

The authors declare that they have no competing interests.

# Supplementary Information

# Realization of one-dimensional electronic flat bands in untwisted moiré superlattice

Yafei Li[1,†], Qing Yuan[1,†], Deping Guo[2,†], Cancan Lou[1], Xingxia Cui[1], Guangqiang Mei[1], Hrvoje Petek[3], Limin Cao[1], Wei Ji[2,*] & Min Feng[1,4,*]

[1]School of Physics and Technology and Key Laboratory of Artificial Micro- and Nano-Structures of Ministry of Education, Wuhan University, Wuhan 430072, China

[2]Beijing Key Laboratory of Optoelectronic Functional Materials & Micro-Nano Devices, Department of Physics, Renmin University of China, Beijing 100872, China

[3]Department of Physics and Astronomy and IQ Initiative, University of Pittsburgh, Pittsburgh, PA 15260, USA

[4]Institute for Advanced Study, Wuhan University, Wuhan 430072, China

†These authors contributed equally to this work.

*Corresponding author. Email: wji@ruc.edu.cn (W.J.); fengmin@whu.edu.cn (M.F.)

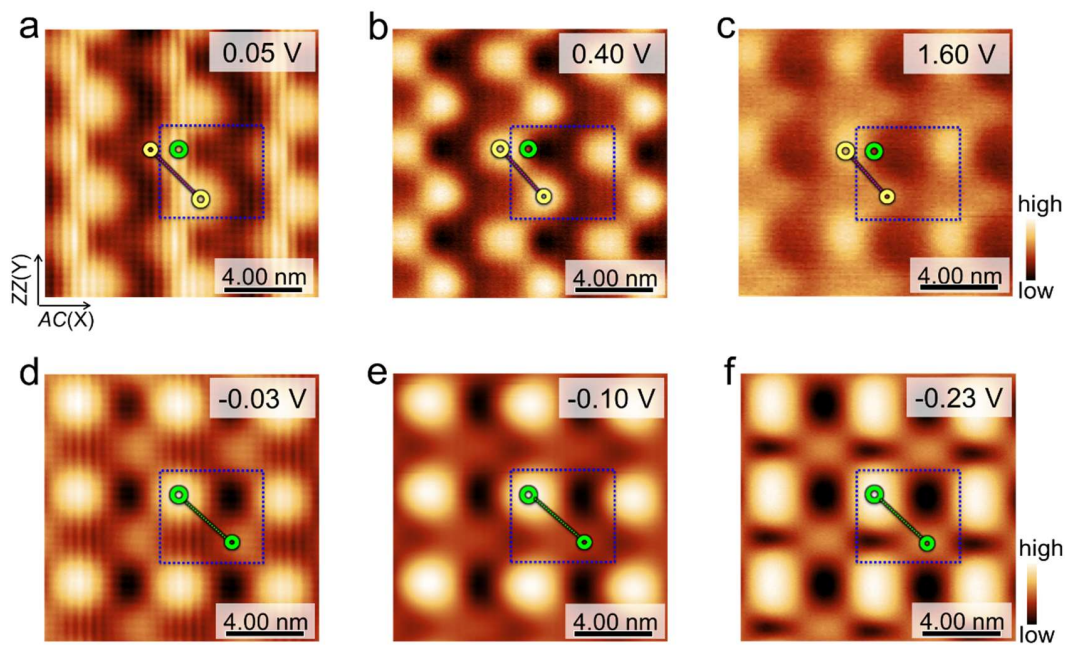

**Fig. S1 Typical STM topographic images of 2-AL Bi(110) superlattice on SnSe(001).**
**a-c** STM images acquired at positive biases exhibiting a zigzag superlattice with the repeating unit marked by yellow circles. **d-f** STM images acquired at negative biases exhibiting a waffle-like superlattice with the repeating unit marked by green circles.

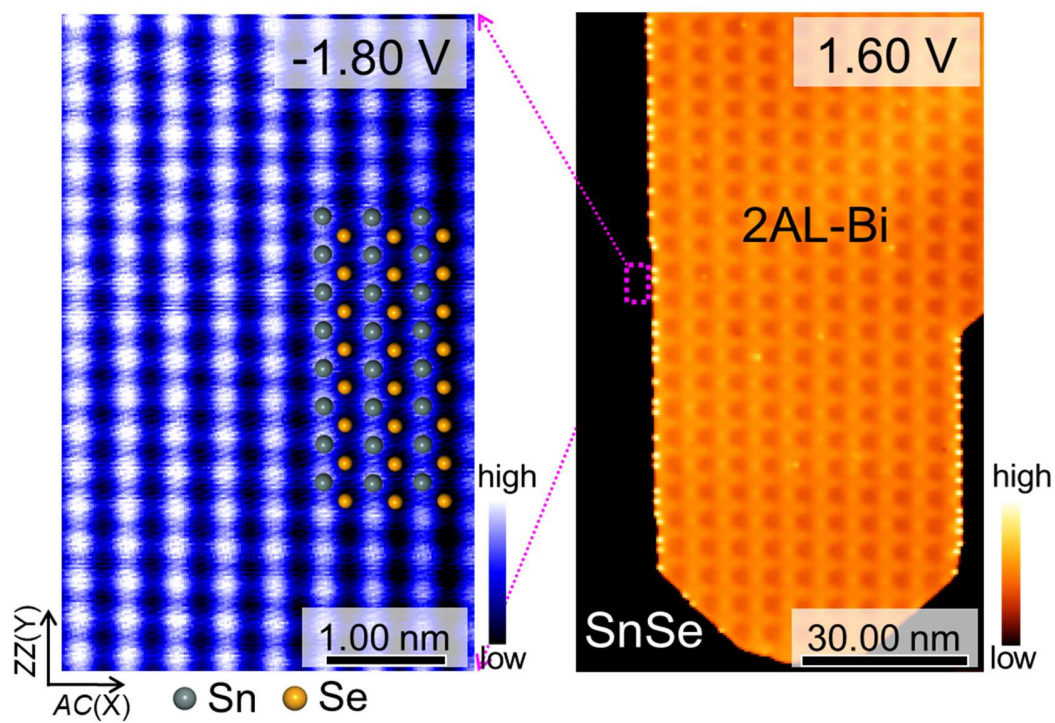

**Fig. S2 STM topographic image of 2-AL Bi(110) islands and SnSe(001)**. The STM contrast shows a waffle-like, rectangular moiré superlattice. The zoomed-in STM image of the pink-rectangle marked region shows the atomic resolution STM image of SnSe acquired just beside Bi islands. The bright spots at the edge of the Bi islands are structural defects which do not influence the alignment of the Bi zigzag chains.

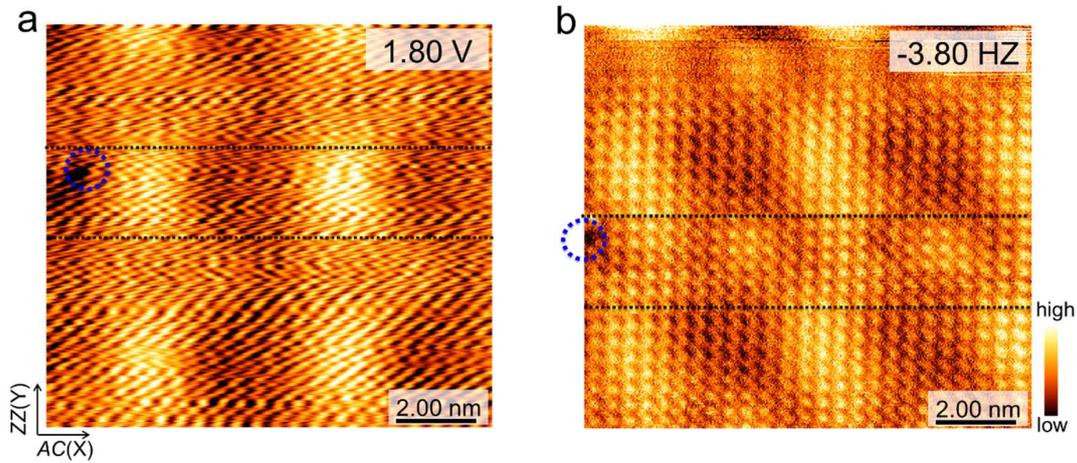

**Fig. S3 Contrast correspondence between the STM and AFM measurements. a** STM image acquired at 1.80 V and 5 pA before the AFM measurement. **b** AFM image obtained at constant Δ*f* mode with a -0.001 V bias, which is acquired just after STM measurement. The blue-dotted circle marks a defect, which serves as the reference for comparison of the STM and AFM images. When changing between the STM and AFM modes, the tip needs to retreat and re-approach, causing a shift between STM and AFM scanning regions. The reference defect, however, enables judging of the contrast correspondence between STM and AFM measurements. Referencing the images based on this defect, it is evident that the highest intensity contrast in STM image does not correspond to highest height contrast in AFM images.

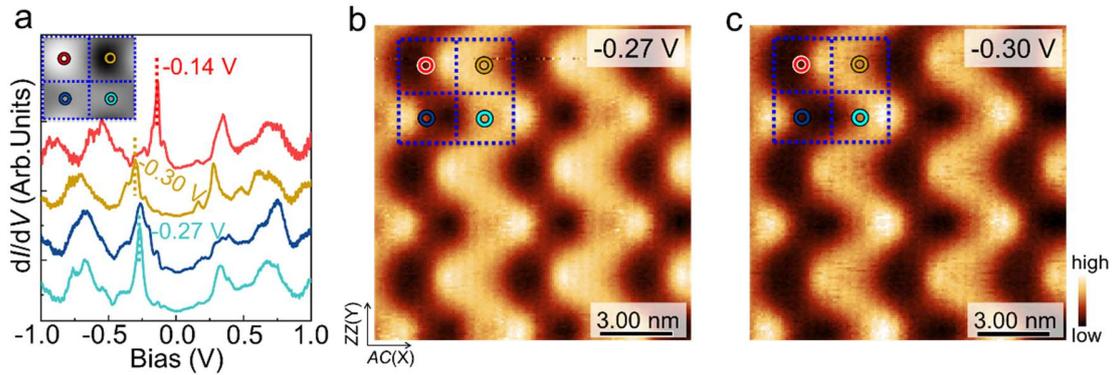

**Fig. S4 Typical STS d$I$/d$V$ spectra and d$I$/d$V$ mappings measured on Bi(110) superlattice on SnSe(001)**. **a** STS d$I$/d$V$ spectra acquired from the centers of Regions I to IV, respectively. Besides a V-dip around zero bias in the spectra, a characteristic feature in the four spectra appears in the occupied state STS regions where a sharp pronounced occupied state STS peak is presented. The sharp peak locates at ~-0.14 V for Region 1, at ~-0.30 V for Region 2. While, the energy level of this peak is ~-0.27 V for the spectra acquired from the center of Regions 3 and 4. **b, c** STS d$I$/d$V$ mappings measured at the energies of -0.27 V and -0.30 V respectively, which show identical morphology and distributions in the real space. We propose they have similar physical origins, except there is somewhat an energy shift between them.

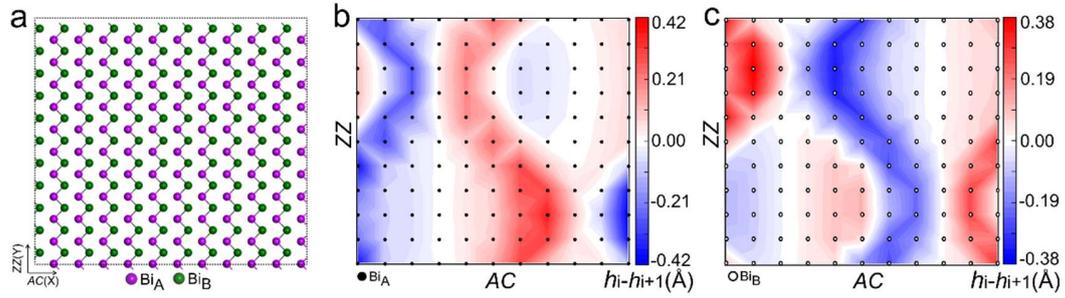

**Fig. S5 Theoretical out-of-plane shear strain distribution in Bi(110) moiré superlattice on SnSe(001). a** Ball and stick model shows the top-view atomic structure of 2-AL Bi(110) layers and the $Bi_A$ and $Bi_B$ atoms in one unit cell are marked. **b** 2D $\Delta h_i$ map of shear strains caused by the $Bi_A$ atoms in the supercell. **c** 2D $\Delta h_i$ map of shear strains caused by $Bi_B$ atoms in the supercell, demonstrating similar shape, but with reversed contrast to (**b**) indicating Bi atom displacement in the opposite direction.

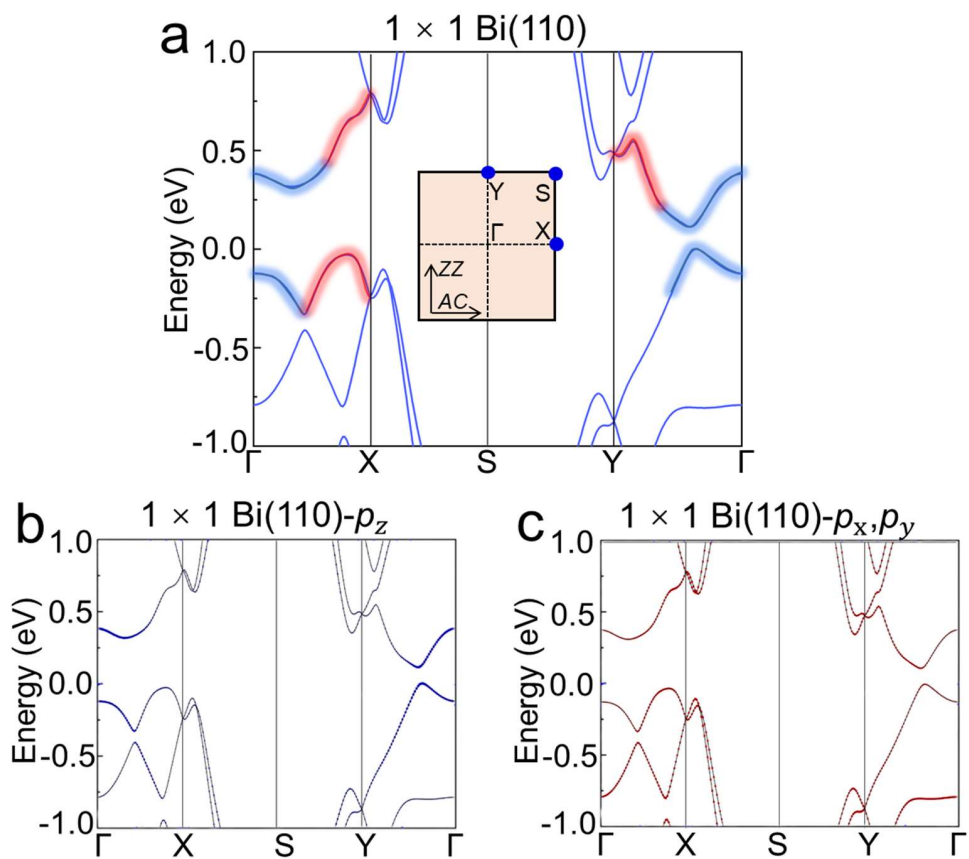

**Fig. S6 DFT calculated band structures of the isolated free 2-AL Bi(110). a** Band structures of the isolated free 2-AL Bi(110) at its equilibrium structure. The blue and red shading highlights the $p_z$ (blue) orbital and $p_x+p_y$ (red) orbitals components, respectively, both in the valence bands (VBs) and conduction bands (CBs). **b, c** Projected contributions from $p_z$ orbital (**b**) and from $p_x+p_y$ orbitals (**c**) for the band structures. The analysis shows that $p_z$ orbitals mainly contribute to the valence and conduction bands around $\Gamma$ point.

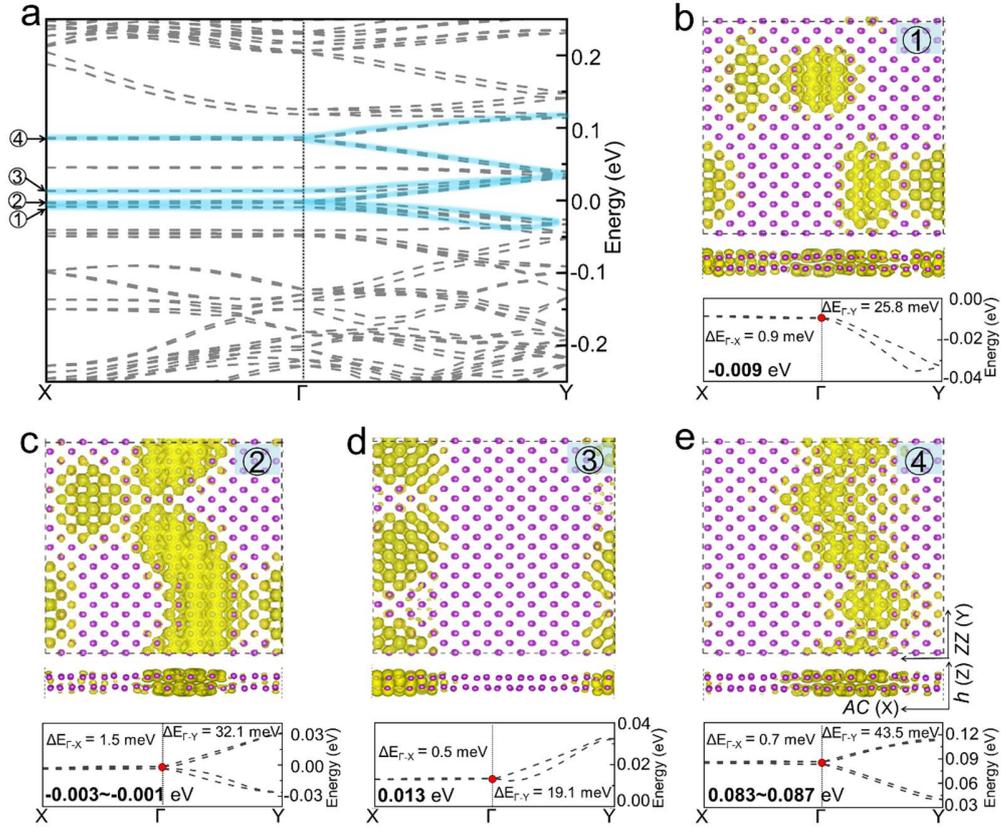

**Fig. S7 The band dispersions along X-Γ-Y path and the corresponding wave function square ($|\varphi|^2$) of the strained 2-AL Bi(110) superlattice without the SnSe substrate**. **a** Calculated band structures along the X-Γ-Y direction. The bands marked by blue shadows and circled numbers are selectively presented in **b-e**. The chosen bands are those contributed by $p_z$ orbitals and those totally flattened along Γ-X direction by the strain. **b-e** Top and side views of $|\varphi|^2$ acquired at Γ point for those highlighted bands in (**a**). Like the two 1D flat bands in Fig. **4e**, **f**, these strain flattened $p_z$ bands (along Γ-X direction) also follow the real space strain distributions. The highlighted bands in (**a**) are zoomed in in the bottom panels in **b-e**, where the red dots mark the energy at the Γ point. The bandwidths of these bands along Γ-X and Γ-Y directions are denoted in meV. It is clear that all of the bands are anisotropic, where the bandwidths along Γ-X direction are smaller than those along the Γ-Y direction.

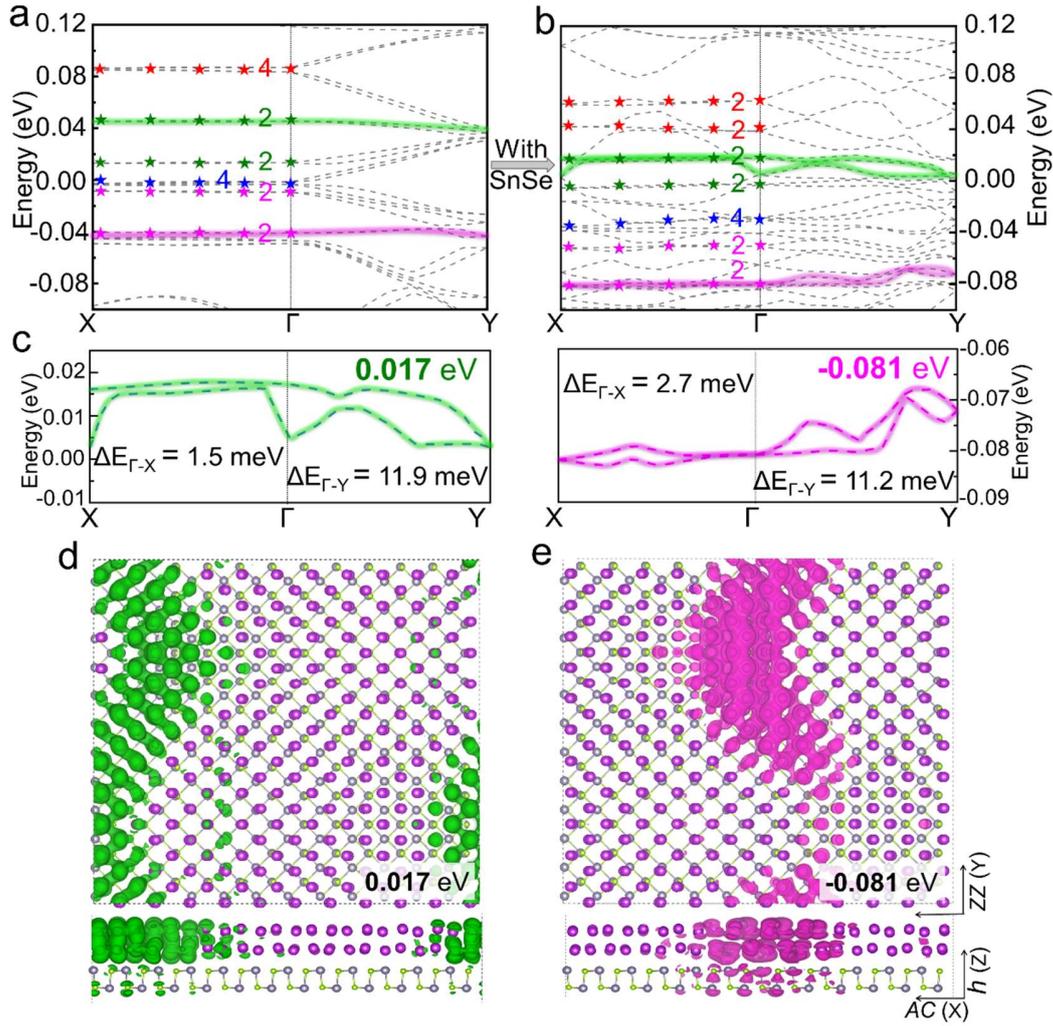

**Fig. S8 The calculated band structures along X-Γ-Y path of the strained 2-AL Bi(110) superlattice without and with SnSe substrate. a** Band structures without the SnSe substrate. **b** Band structures with the SnSe substrate. The stars of different colors mark the strain flattened $p_z$ band along Γ-X direction. The colored numbers on the bands indicate the band degeneracy of the corresponding bands. Comparison of (**a**) and (**b**) shows that, the flat bands predicted without SnSe (**a**) can still be resolved in (**b**) when the SnSe substrate is included. These bands retain the flat dispersion character along the Γ-X direction, however, a rigid energy down-shift is predicted in (**b**). Considering the two 1D flat bands at 0.045 eV and -0.041 eV in (**a**), they keep the 1D flat band character in (**b**) without noticeable changes of the dispersions both along Γ-X and Γ-Y directions. **c**

Amplified images showing the bandwidth along X-Γ-Y path for the two highlighted 1D electronic flat bands in (**b**). The bandwidths of the two 1D electronic flat bands, along Γ-X and Γ-Y directions, do not change much as those in Fig. 4**c**, **d** where there is no substrate. **d, e** $|\varphi|^2$ of the two 1D flat bands at Γ point also follow the strain distributions, exhibit the same real space distributions as those of the States A and B in Fig. 4**e** and **f**, where the SnSe is not included in the calculations. The results reveal that including the substrate does not perturb the electronic structures of the strain-induced 1D flat bands.

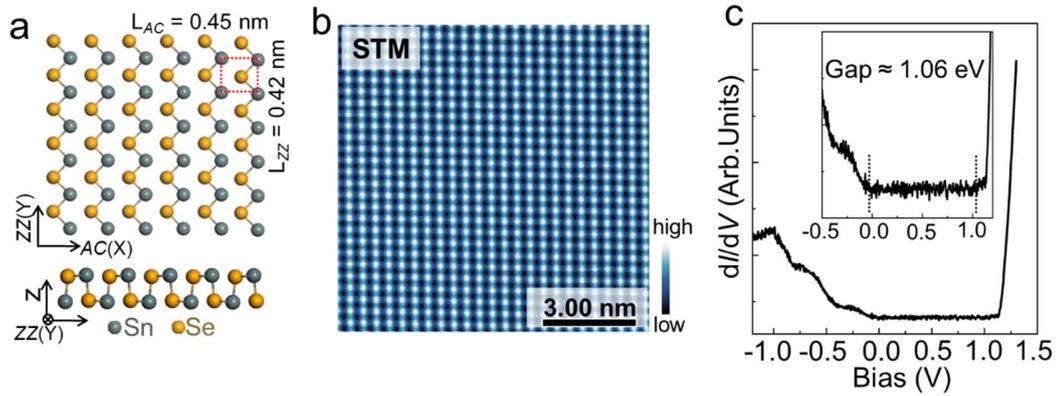

**Fig. S9 SnSe single crystals**. **a** Ball and stick model of the atomic structure of SnSe. The upper and lower panels show the top and side views of SnSe(001), which illustrate the armchair and zigzag structures of SnSe(001), respectively. The atomic unit cell is presented with experimental data. **b** Atomic resolution STM image reveals the atomic structure of the SnSe(001) surface. Because of the buckled structure of SnSe(001) surface, normally only Sn atoms can be resolved clearly. The atomic structure of SnSe(001) captured in our STM image is consistent with previous reports[1-3]. **c** The d$I$/d$V$ spectrum of SnSe, indicating its p-type semiconductor characteristics.

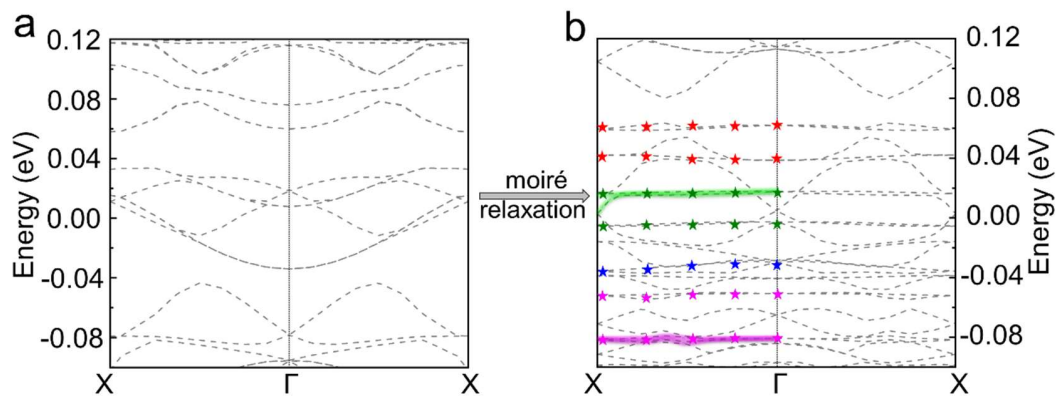

**Fig. S10 DFT calculated band structures of 2-AL Bi(110) on SnSe without (a) and with (b) the moiré relaxation.** It can be clearly seen that in (**a**), without atomic relaxation, there are no flat bands. As shown in (**b**), however, the flat electronic bands arise from atomic relaxation, which correspond to those marked in Fig. S**8**.